\documentclass[prl, amsfonts, twocolumn, nofootinbib, showpacs]{revtex4}
\usepackage{graphicx, epsfig}

\def\sm{{\tilde{m}}}
\begin{document}
\title{A new family dependent interaction in Tevatron top dilepton candidate events ?}

\author{J.~L. Diaz-Cruz$^1$}
\affiliation{$^1$ Cuerpo Acad\'emico de Particulas, Campos y Relatividad FCFM,
 		  BUAP, Ap.Postal 1364 Puebla Pue.~72000, Mexico}
\author{C.~E. Pagliarone$^2$}
\affiliation{$^2$ Universit\'a degli Studi di Cassino \& I.N.F.N. Pisa,\\
                  via F. Buonarroti n. 2 - 56127 Pisa, Italy.}

\begin{abstract}
New family dependent fermionic interactions have been
conjectured in several extensions of the Standard Model that range 
from Supersymmetry to composite theory up to flavor interactions. 
Strong constraints on these theoretical scenarios can be derived  
from light fermion phenomenology and from $B$-mesons studies. 
Corresponding constrains on the top quark sector are, on the other hand, rather week.
Tevatron data, on top quark pair production and decay in dilepton channel, 
may suggest some deviation from the Standard Model expectations. 
Such a deviation can be successfully re-interpreted in terms of an exotic 
top decay that can arise in several theories beyond the Standard Model.
Further investigations at present and future colliders will provide 
crucial tests on the models under discussion.
\end{abstract}
\pacs{14.65.Ha,12.60.-i}
\maketitle

\section{I.   $\,\,$Introduction}  
\vspace{-0.55cm} 
After the discovery of the top quark at Fermilab Tevatron Collider~\cite{cdf_top,d0_top},
experimental attention has been turned on the examination of its 
production mechanisms and decay properties. 
Within the Standard Model (SM), the top quark production cross section is 
evaluated with an uncertainty that is of the order of $\sim 15\%$ and top quarks
are assumed to decay to a $W$ boson and a $b$ quark almost $100\%$ of the time.
In the fermion sector of the SM, plenty of experimental data, have proved that 
interactions of quarks and leptons, with gauge bosons, are correctly 
described by the $SU(3)_c\times SU(2)_L\times U(1)_Y$ gauge theory~\cite{revsm}. 
At tree-level, SM neutral interactions are diagonal in flavor space, 
however, due to the generation mixing, that appears in the 
charged currents, flavor changing neutral currents (FCNC) 
can arise at loop level.
The fact that FCNC $B$-meson decays have been already detected, at
rates consistent with the SM~\cite{PDG}, represents a great 
success for the model itself.
Current bounds, on rare decays of light quarks and leptons, 
impose further constraints on new sources of FCNC transitions. 
However, SM predictions for related processes, involving  
top quarks, are strongly suppressed, although corresponding  
experimental bounds are rather weak.  

Due to its exceedingly heavy mass, it is reasonable to expect the top to
be more related to new physics than other fermions.
Top quark physics may then serve as an important window for probing physics beyond
the SM. 
A significant deviation from SM prediction would then indicate either a 
novel production mechanism or a novel decay channel that is the subject of
the present paper.
As matter of fact, it could be even possible that such effects may be 
observed at present Tevatron collider experiments (CDF-II, D0)
or in the next generation collider experiments at LHC~\cite{tt-rev} 
or later on at the next Linear Collider~\cite{tt-LC}.
FCNC top quark decays, such as
$t\to c \gamma$, $t\to c g$, $t\to c Z$ and $t\to c \phi$ 
have been studied, for some time, as possible signals of   
physics beyond the SM~\cite{raretop}.
The extended models, that we are interested to discuss, in the 
present article, include both top-charm (TC) transitions and 
$\mu-\tau$ violating (MTV) interactions, whose combined effect 
can induce top quark decays such as $t \rightarrow c \tau \mu$ or 
$t\rightarrow u \tau \mu$. Such processes may arise either through a scalar 
bosons, with family-dependent couplings, or by direct contact interactions.
We shall refer to $t\rightarrow c(u) \tau \mu$ as double flavor-violating (DFV)
top decays.

\vspace{-0.4cm} 
\section{II.   $\,\,$Theoretical Motivations}
\vspace{-0.4cm} 
Past and present Tevatron top analysis have been performed, always 
assuming, a dominant SM branching ratio ($Br(t \to bW) \simeq 1$).
The possibility that exotic top quark decay modes could affect the 
dilepton top quark events ($p \bar p \rightarrow t \bar t \rightarrow 
W^{+}b W^{-}\bar b \rightarrow b\ell^{+} \nu_{\ell} \bar b \ell^{-} \bar{\nu_{\ell}}$)
have not been considered yet.
In this paper, we are interested in studying how the dilepton signature,
could be affected 
when one includes additional top decay channels such as: $t \rightarrow c \tau \mu$ or
$t \rightarrow u \tau \mu$.
We will argue later that such an effect could have already shown up  
and may be available in the top  
quark data published by the CDF collaboration~\cite{CDFdil1},\cite{CDFdil2}.  

Since in the SM, the branching ratio of top decaying into  
$c(u) \mu \tau$ is extremely suppressed~\cite{SM-tch,SM-MTV},  
then the study of this decay channels may be an excellent way 
for searching for physics beyond the SM. 
In the next sections we will discuss the expected branching ratios arising  
in several and well motivated extensions of the SM,  
which have been proposed to tackle some open issues.

Among these theories, we shall discuss first Supersymmetry (SUSY), which is  
studied mainly in connection with a possible solution of the hierarchy  
problem~\cite{susyrev}.  Its minimal incarnation, the Minimal  
Supersymmetric extension of the SM (MSSM), has well know attractive features, 
such as allowing gauge coupling unification, and explaining
electroweak symmetry breaking (EWSB) as a radiative effect, driven 
by the large top quark Yukawa coupling.  
In addition  SUSY also predicts tau-bottom Yukawa unification 
and provides a dark matter candidate.  
Within the context of SUSY models, it is possible to induce FCNC Higgs 
interactions, which could mediate the MTV top decay $t \to c \mu \tau$ 

The resulting structure of the Higgs-Yukawa sector in MSSM, corrected by loop effects,
actually shares similarities with the general version of the two-Higgs doublet 
model (THDM-III), whose predictions for DFV top decays will be 
discussed, for comparison, with other models.
The Yukawa structure of the THDM-III relies on the assumption of a
specific texture, for fermion mass matrices, which makes possible to 
derive a pattern for TC and MTV interactions, which is possible to constrain 
with low-energy physics experiments, making also available predictions 
at higher energy~\cite{ldc-rnp-ars}. 
A particular successful ansazt, for the
THDM-III, consists in assuming a four-texture for the mass matrices, 
which is able to reproduce quark and charged lepton masses, as well as 
the CKM mixing matrix. 
This approach can be considered as an starting
point to address one of the most challenging problems of the SM, 
the flavor problem, namely, how to explain the observed pattern 
of quark and lepton masses and their mixing angles.

Along the lines of flavor physics, we are also going to discuss models 
that attempt to explain the structure of the fermion mass 
matrices by an approximate flavor symmetry. 
Models, with either Abelian or non-Abelian flavor symmetry, have 
been proposed in literature~\cite{flavormods}. 
The scale, associated with these models, could range from 
TeV up to the Grand Unification scale (GUT scale).  
Flavor Models, with a low scale flavor symmetry breaking, offer a  possibility  
to observe direct effects of  flavor physics, for 
instance through the detection of scalar particle 
associated with the mechanism of flavor symmetry breaking, 
the so called {\it{Flavon Fields}}. 
On the other hand, when the flavor symmetry is a global-type symmetry, 
its breaking leaves a light pseudo-goldstone bosons, 
the {\it{ familons}} that could provide
detectable signatures too~\cite{familonmods}. 
The effects of the flavons or familons could even be transmitted to 
the EWK  Higgs sector, through mixing \cite{flavons}, 
and could induce exotic Higgs bosons 
and top quark decays.

Finally we shall also consider strongly interacting scenarios~\cite{strongtop}.
These models are interesting as they offer some clue about the nature 
of EWSB.  As matter of fact, such models often predict new physics to manifest itself 
in the top quark sector mainly because of the large top-quark mass.
They can also arise because of the presence of additional 
substructure layers, which could manifest themselves first for the 
heaviest SM fermion. 
Strong interaction scenarios could then induce exotic DFV top quark decays too.

\vspace{-0.8cm} 
\section{III.  $\,$TC  $\,$and  $\,$ MTV  $\,$Models} 
\vspace{-0.3cm} 
\subsection{A.  $\,$Overview} 
\vspace{-0.3cm} 
In order to search for effects of a new family-dependent 
interaction (FDI), in top quark decays, we will start considering 
first  the case when such interaction is mediated by a new boson ($S^0$).
The case, where such decays occurs through an unsuppressed contact  
interaction, will be discussed separately.
The interactions of $S^0$ is here described by an effective  
lagrangian, that parametrizes family-dependent couplings for quarks   
and leptons, $q,l_{L,R}$, in terms of form factors: $F^{q,l}_{L,R}$.
For the top-charm and tau-muon couplings the lagrangian takes the 
following form:
\begin{equation} 
{\cal{L}}=  S^0 \bar{t} [ F^{u}_L P_L +   F^{u}_R P_R]c + 
            S^0 \bar{\tau} [ F^{l}_L P_L +   F^{l}_R P_R]\mu +h.c. 
\end{equation} 
In the present case, the top quark decay $t\to c \mu \tau$ 
will proceed through the  $S^0$ intermediate state, 
namely $t\to cS^0$ and then $S^0 \to \mu\tau$.  

\vspace{-0.3cm} 
\subsection{B.  $\,$TC  $\,$ and  $\,$ MTV  $\,$ in Minimal SUSY Models} 
\vspace{-0.3cm} 
 
Within the context of the MSSM, we shall consider that the  
double flavor-violating (DFV) top quark decays $t \rightarrow c \mu \tau$ 
are mediated through TC and MTV Higgs interactions, which arise 
through loops involving gauginos and sfermions. 
In particular, lepton flavor violation (LFV), experimentally observed 
in the neutrino sector~\cite{LFVneutrino}, could be transmitted from the SUSY breaking 
sector to sfermions, and then from the sfermions to the Higgs sector.
In this way the tree-level type-II Higgs sector of MSSM becomes 
a type-III THDM.   
Thus, LFV Higgs interactions in MSSM, which induce decays such as  
$\phi_i \rightarrow \mu \tau$ ($\phi_i=h,H,A$), are produced by loops involving 
sleptons and  charginos or neutralinos~\cite{myLFVH}.
Moreover, because SUSY GUT's relate quarks and leptons, it is possible 
that  the large mixing, observed in the neutrino sector, implies 
a large mixing in the up-type squarks, which in turn, could also  
induce the FCNC top decay $t \rightarrow c h^{0}$ ~\cite{ourTCH}.  
By combining these effects it is possible to have 
exotic top quark decay $t \rightarrow c h^{0}$, followed by
$h^{0}\to \mu \tau$. 

We shall consider a minimal SUSY-FCNC schemes, where a large mixing 
appears in the slepton and squark trilinear $A$-terms \cite{ourTCH}. 
In particular 
we will consider that the elements of $A$-terms satisfy 
the conditions: $A_{33}=A_0$,$A^u_{23}= x_u A^u_0$, $A^l_{23}= x_l A^l_0$, 
where $x_{u,l}$ are  
unknown  $O(1)$ parameters, $A_0$ denotes an universal SUSY breaking 
trilinear term and the remaining $A$-entries are assumed to vanish.  
After diagonalizing the resulting sfermion 
mass matrices, and writing the interactions in terms of mass 
eigenstates, one can evaluate the TC/MTV top and Higgs decays. 
The branching ratios will be sensitive to the 
values of $x_{u,l}$, since these parameters affects both the sfermion  
mass-eigenvalues (that enter the loop integrals) and the  
TC/MTV couplings in the $\tilde{t}-\tilde{c}$ 
and $\tilde{\mu}-\tilde{\tau}$ sectors. 

\vspace{-0.3cm} 
\subsection{C.  $\,$TC and MTV interactions in the THDM-III} 
\vspace{-0.3cm} 
 
The THDM-III, based on the assumption of a  specific four texture 
for the fermion mass matrices, has been extensively discussed in
Ref.~\cite{ldc-rnp-ars}. 
In that paper we derived the resulting pattern for TC and MTV Higgs 
interactions, which generalizes the Cheng-Sher ansazt~\cite{chengsher}; 
namely the phases allowed by the hermitian four-texture ansazt were also 
included in the analysis. 
We derived also constraints on the model parameters,
by comparing low-energy physics processes (mainly: $\mu \to e+\gamma$, 
$\tau \to 3\mu$, $e-\mu$ conversion and rare $B$-decays). This allowed us
to make predictions on the detectability of MTV Higgs decays at present
and higher-energy colliders~\cite{cotti}. 

In this model, the couplings, of the lightest Higgs boson ($h^0$)
with fermions of different flavor, can be written as follows:
\begin{equation}
{\cal{L}}_{hf_if_j}=  \frac{\sqrt{m_i m_j}}{v} \tilde{\chi}_{ij}\eta_h 
                        \bar{f}_i f_j h^0
\end{equation} 
where $m_{i,j}$ corresponds to the masses of the $i$ and $j$ fermions, 
$v$ denotes the scale of EWSB and $\tilde{\chi}_{ij}$
are unknown $O(1)$ coefficients.
The factor $\eta_h$ depends on the fermion type. For $u$-type
quarks its value is: $\eta_h=-\cos(\beta-\alpha)/2\sqrt{2} \sin\beta$,
while for $d$-type quarks and leptons:  
$\eta_h= \cos(\beta-\alpha)/2\sqrt{2} \cos\beta$, where
$\alpha$ denotes the mixing angle in the $CP$-even Higgs sector. 
Therefore, the form factors for the couplings $h^0 f_i \bar{f}_j$ 
are: $F_L=F_R= \sqrt{m_i m_j} \chi^{u,l}_{23} \eta_h/v$.

\vspace{-0.3cm} 
\subsection{D.  $\,$Flavored bosons and TC/MTV top decays} 
\vspace{-0.3cm} 
 
Models  with a low flavor scale, which offer the possibility  
to observe direct  flavor physics effects, could also 
predict large rates for the decay: $t\to c\mu\tau$. 
One way to mediate this decay can be provided by  
the scalar particle responsible for the flavor symmetry  
breaking, the so called {\it{ flavon fields}} ($\Phi^0_F$).  
In this case,  we shall consider a model that contain 
a scalar particle that couples to fermions of 
different generations. A convenient parametrization,  
for its couplings, is provided by the Cheng-Sher ansazt~\cite{chengsher}. 
In order to maximize the effects, in the following analysis,
we shall assume a scenario where the flavon field couples 
predominantly to fermions of different generations.
Such a pattern can arise when flavor symmetry is
non-abelian~\cite{familonmods}. 
Then, we  assume that the mass-eigenstate 
$S^0$ results form the mixing between a flavon field $\Phi^0_F$ 
and a Higgs boson $H^0$.  
The form factors for the coupling $S^0 f_i \bar{f}_j$ 
are:  
\begin{equation} 
F_L=F_R=\sqrt{m_i m_j} \frac{\chi_{ij}}{v} 
[(1-\delta_{ij})\sin\xi+\cos\xi\delta_{ij}] 
\end{equation} 
where $m_{i,j}$ correspond to the masses of the fermion $i$ and $j$, 
$v$ denotes the v.e.v., that breaks the electroweak symmetry, 
and  $\xi$ is the angle that parametrizes the mixing between the 
Higgs and the flavon fields, in the scalar sector.
The factor $\chi_{ij}$ is
again an $O(1)$ unknown coefficient.

\vspace{-0.3cm}
\subsection{E. $\,$Strong Dynamics} 
\vspace{-0.3cm} 

Furthermore, because the top quark is the heaviest SM fermion, 
it could offer some clue about the nature of EWSB, or could even be 
a composite state. In any case, it is likely that 
strong dynamics could induce rare top quark decays too. 
In this case, the strong  dynamics could be associated with models 
of composite quarks and leptons and it is possible to have  
contact interactions, that can also mediate exotic 
transitions. The contact interaction for $tc\tau\mu$, including 
only a scalar or a pseudoscalar coupling, is given by: 
\begin{equation} 
{\cal{L}}=\frac{c_{23}}{\Lambda^2}\, \bar{t}\,\Gamma_{S}\, c  
                       \bar{\tau}\,\Gamma^{'}_{S} \, \mu +h.c. 
\end{equation} 
where $c_{23}$ is an unknown  coefficient that has been calculated 
to be about $m^2_t/96 \pi$~\cite{tctaumu}
and $\Gamma_s=1 (\gamma_5)$ for scalar (pseudoscalar) couplings. 

\vspace{-0.2cm} 
\section{IV. $\,$ Results on DFV top $\,\mathbf{Br(t\to c \tau \mu)}$} 
\vspace{-0.3cm}  

We are now interested in the evaluation of the double-flavor-violating
top decay Branching Ratios, $Br(t \to c\mu\tau)$, for all the models 
discussed above.
When this decays are mediated by the scalar resonance $S^0$, we can 
factorize the branching ratio as follow:  
\begin{equation} 
Br(t \to c \tau \mu)= Br(t\to c S^0)\times Br(S^0\to \mu \tau)
\end{equation} 
where $Br(t \to c S^0)= \Gamma(t \to c S^0)/ \Gamma_t$. For the total 
top decay width, we will use the approximation:
$\Gamma_t = \Gamma(t \to bW^+)_{\rm SM}\,$ and
$\,Br(S^0 \to \mu\tau)= \Gamma(S^0 \to \mu \tau)/\Gamma_S$.
At this point the total width of the  scalar will depend only from the model.
The decay widths for $t \to c S^0$ and $S^0 \to \mu \tau$ 
are then written in terms of form factors ($F_{L,R}$) as follows: 
\begin{eqnarray}  
\Gamma (t \to c S^0_i) &=& \frac{m_t}{16 \pi} (1-\frac{m^2_h}{m^2_t})^{1/2}     
                   [ |F^t_L|^2 + |F^t_R|^2  ] \\  
\Gamma (S^0\tau \mu) &=& \frac{M_S}{8\pi}[ |F^l_L|^2 + |F^l_R|^2  ] 
\end{eqnarray}  

\noindent
We find the following results for each of the model considered. 

\vspace{0.1cm} 
{\bf{ 1. MSSM}}. In the case of the MSSM,   
the  DFV decay $t\to c S^{0}$ followed by $S^{0} \to \mu \tau$ 
is always kinematically allowed for $S^0=h^0$.
Then, as shown in table 1, the decay branching ratio 
$Br(t\to ch^0)$ can be as large as $10^{-3} - 10^{-4}$, 
over a large part of the SUSY parameter  
space, where the mass of the lightest Higgs boson $h^0$  
is around $110\div 130 \,GeV$~\cite{ourTCH}. 
For a gluino mass: $M_{\tilde g} =500$ GeV, 
pseudo-scalar Higgs mass: $m_A=300$ GeV, SUSY parameters
$(\sm,\mu,A_u)=(0.6,0.3,1.5)$ TeV, $x_u=0.9$, $\tan\beta=5\, (10)$, 
one finds:
$Br(t\to c h^0) = 4.0 \times 10^{-4} \, (10^{-3}) $.
Similarly, one can evaluate the decay $h \to \mu \tau$, and 
finds a branching ratio of the order of
$10^{-4}\div 10^{-5}$~\cite{myLFVH}. In fact, for a Bino-mass
$m_{\tilde B}=600$ GeV, $\,(\sm,\mu,A_l)=(0.6,0.3,0.3)$ TeV,
$x_l=0.9$ and $\tan\beta=5 (10)$ we get: 
$Br(h^0 \to \mu\tau)= 4.4 \times 10^{-4} (1.2\times 10^{-4})$.
\vspace*{-1.4mm} 
\begin{center}
\begin{table}[t!]
\begin{ruledtabular} 
\begin{tabular}{|c c | l r | l r | l r|} 
$\mathbf{M_{A}}$ & $\mathbf{M_{\tilde g}}$ &  \multicolumn{2}{c}{$\mathbf{tan\beta=5}$} 
&  \multicolumn{2}{c}{$\mathbf{tan\beta=20}$} &  \multicolumn{2}{c|}{$\mathbf{tan\beta = 50}$}\\ 
 & &  \multicolumn{2}{c}{$(\times 10^{-5})$} 
&  \multicolumn{2}{c}{$(\times 10^{-5})$} 
&  \multicolumn{2}{c|}{$(\times 10^{-5})$}\\
\hline 
150 &   $250$   & $1.1$  & $57$   &  $1.9$  & $200$  &  $2.1$  &  $260$\\
\hline 
150 &  $500$   & $0.9$  & $33$   &  $1.4$  &  $95$  &  $1.6$  &  $120$\\ 
\hline 
300 &  $250$   & $1.4$  & $69$   &  $2.0$  & $210$  &  $2.1$  &  $260$\\
\hline 
300 &  $500$   & $1.1$  & $40$   &  $1.5$  &  $99$  &  $1.6$  &  $120$\\
\hline 
600 &  $250$   & $1.4$  & $70$   &  $2.0$  & $210$  &  $2.1$  & $270$\\
\hline 
600 & $500$    & $1.1$  & $41$   &  $1.5$  & $100$  &  $1.6$  & $120$\\
\end{tabular} 
\vspace{-0.35cm}
\caption{ $Br(t\rightarrow c\,h^0)$ for a sample set of $A1$-type 
inputs with $(\sm0,\mu,A^u_0)=$ $(0.6$, $0.3$, $1.5)$\,$TeV$. 
The two numbers in each entry correspond to 
$x = (0.5,\,0.9)$, respectively. $m_A$ and $m_{\tilde g}$ 
are expressed in $GeV$.}
\end{ruledtabular}
\vspace{-0.5cm}
\end{table}
\end{center} 
\vspace{-0.4cm}
Therefore the branching ratio for the 
DFV top quark decay $t\to c \mu \tau$ reaches a value of the order 
$2\times 10^{-7}$ for $\tan\beta=5 (10)$, which is well bellow present
top analysis sensitivity.

\vspace{0.1cm} 
{\bf{2. The THDM-III}}.
For the THDM-III we have that low energy physics implies
that $\chi^l_{23}$ is constrained to be of order $10^{-1}$,
while $\chi^u_{23}$ remains essentially unconstrained.
Therefore, we can take it of order one.
For $m_h=120$ GeV, and $5 < \tan\beta < 50$,
$\alpha=\beta-\pi/3$, it is possible to obtain: $\chi^l_{23}=0.1$, and
$Br(h^0 \to \mu\tau)$ only gets up to about $2\times 10^{-5}$, while for 
the TC top decay we find: $Br(t\to ch^0) \simeq 0.1$. Then,
for DFV top decay, we calculate:
$Br(t\to c\mu\tau)=2\times 10^{-6}$.
It should be pointed out that low values for $Br(h^0 \to \mu\tau)$
are obtained in this model, because its flavor structure forces the
coefficients to satisfy the condition: $\chi^l_{23}=\chi^l_{13}$, which then
allows to use the rare decay $\mu\to e \gamma$ to constrain the
parameter. 
When this relation no longer holds, 
it is possible to have: $\chi^l_{23} \simeq 1$, and then
this makes possible to obtain: $Br(h^0 \to \mu\tau)\simeq 4 \times 10^{-2}$,
and therefore: $Br(t\to c\mu\tau)=4\times 10^{-3}$.

\vspace{0.1cm}  
{\bf{3. Flavon-Higgs Mixing Model}}. 
The result for the branching ratio will depend
on the mixing angle $\xi$, introduced to  
parametrize the mixing between the Higgs and flavon fields. 
We shall assume that the lightest mass-eigenstate, which is denoted by
$S^0$, has mass in the intermediate range: $m_Z < m_S < m_t -m_c$.  
Then, in order to compute its branching ratio, we shall approximate it as:
$Br(S^0 \to \mu \tau) = \Gamma(S^0 \to \tau \mu)/ \Gamma(S^0 \to b\bar{b})$. 
For a reasonable range of parameters, $\sin\xi =0.9$, $m_S=120$ GeV, 
we find that $Br(t \to c S^0) \simeq 0.1$,  
$Br(S^0 \to \mu\tau)\simeq 0.4$, therefore  
$Br(t \to c\mu\tau) \simeq 0.04$. 
As the parameters $\chi^l_{23}$ and $\xi$ are essentially unconstrained, 
larger values for $Br(t \to c\mu\tau)$ around 0.1, are 
also achievable in this model.

\vspace{0.1cm}  
{\bf{4. Strongly Interacting Model}}. 
In this scenario, there is a contact interaction that 
induces directly the decay $t\to c\mu\tau$, with a partial width 
given by: 
\begin{equation}  
\Gamma (t \to c \tau \mu) = \frac{1}{96 \pi} \frac{m^5_t}{\Lambda}  
\end{equation}  
Since there are no constraints on such operators, one can even have 
$\Lambda\simeq m_t$, which then gives a branching fraction
for DFV top decay of order of $\simeq 0.1 \div 0.2$.
A summary of all the results found is shown in Table~II. 

\vspace{-0.5cm} 
\section{IV.   $\,$MTV  $\,$interactions  $\,$and  $\,$$\mathbf{ 
t \bar t \rightarrow b\ell^{+} \nu_{\ell} \bar b \ell^{-} \bar{\nu_{\ell}} }$}

\vspace{-0.2cm} 
\subsection{A.   $\,$Overview}
\vspace{-0.3cm} 
Within the SM, a top, with a mass above $Wb$ threshold, is predicted  
to have a decay width dominated by the two-body process: $t \rightarrow W b\,\,$ 
($Br\simeq 0.998$)~\cite{PDG}.
A $t \bar t$ pair will then decay via one of the following channels, 
classified according to the final state:  
%
$i)$ {\it Dilepton}, where both $W$ decays are leptonic ($\ell_{W}$), 
with 2 jets arising from the two $b$-quark hadronization and 
missing transverse energy ($\not\!\!E_{\rm T}$) coming 
from the undetected neutrinos~\cite{CDFdef}:
$Br(e,e)=Br(\mu,\mu)=Br(\tau,\tau) \simeq 1/81$, $Br(e,\mu)=Br(e,\tau)=Br(\mu,\tau) \simeq 2/81$;   
%
$ii)$ {\it Lepton+jets}, where one $W$ decays leptonically and the other one  
into quarks, with 4 jets and $\not\!\!E_{\rm T}$: 
$Br(e+jets)=Br(\mu+jets)=Br(\tau+jets) \simeq 12/81$;
%
$iii)$ {\it All-hadronic}, where both the $W$'s decay into quarks with 6 
jets and no associated $\not\!\!E_{\rm T}$: 
$Br(jets) \simeq 36/81$.
A $\tau$, coming from $W$, will then decay either in $e\overline{\nu}_{e}\nu_{\tau}$ or 
$\mu\overline{\nu}_{\mu}\nu_{\tau}$ ($\tau_{\ell}$) 
or in a narrow jet ($\tau_{h}$), with branching ratios that are: 
$\Gamma^{\tau}(\mu^{-} \overline{\nu}_{\mu} \nu_{\tau})/\Gamma^{\tau}_{total}= 0.1736 \pm 0.006$,
$\Gamma^{\tau}(e^{-} \overline{\nu}_{e} \nu_{\tau})/\Gamma^{\tau}_{total}= 0.1784 \pm 0.006$
and
$\Gamma^{\tau}(\tau \to jet)/\Gamma^{\tau}_{total} \simeq 0.64$~\cite{PDG}.
Then $\tau$'s coming from $t \bar t$ decay will give rise to the following
final states:
$\not\!\!E_{\rm T} \tau_{h} \tau_{h} b \bar{b}$, 
$\not\!\!E_{\rm T} \tau_{h} \tau_{\ell} b \bar{b}$, 
$\not\!\!E_{\rm T} \tau_{\ell^{'}} \tau_{\ell} b \bar{b}$,
$\not\!\!E_{\rm T} \ell_{W} \tau_{h} b \bar{b}$ and 
$\not\!\!E_{\rm T} \ell_{W} \tau_{\ell} b \bar{b}$.
Further on, we will call dilepton only the following 
final states: $ee+X$, $e\mu+X$ or $\mu\mu+X$.

In both CDF Run 1 and Run 2 $t \bar{t}$ dilepton analysis,
events were selected by requiring two opposite sign (OS)  
high-$P_{T}$ leptons with $P_{T}^{\ell}>$ $20$ $GeV/c$ 
($|\eta^{\ell}|<$ $1.0$), at least two jets with measured 
$E_{T}^{jet}>$ $10$ $GeV$ ($|\eta^{jet}|<$ $2.0$), 
and $\not\!\!E_{\rm T}>$ $25$ $GeV$.
To ensure that the $\not\!\!E_{\rm T}$ is not due to 
mismeasurements, events were also required to have
$\not\!\!E_{\rm T}>$ $50$ $GeV$, if the azimuthal
angle between the direction of $\overrightarrow{\not\!\!E_{\rm T}}$
and the nearest lepton or jet was less than $20^{0}$.
In order to enhance the purity of the sample a further cut 
was applied; $H_{T}$, the scalar sum
of $P_{T}^{\ell}$, $E^{jet}_{T}$ and $\not\!\!E_{\rm T}$,
was required to exceed $170$ ($200$) $GeV$ in Run 1 (Run 2) analysis~\cite{CDFdil1,CDFdil2}.
Using $109 \pm 7$ $pb^{-1}$ of Run 1 data, CDF found, 
in total, $8$ events: $0$ $ee$, $1$ $\mu \mu$ and $7$ $e \mu$
when, for example, $3.2$ events was expected (top plus SM background~\cite{backgrounds})
in the $e \mu$ channel~\cite{CDFdil1}.
In Run 2, using $193$ $pb^{-1}$ of data,
CDF-II observed $13$ events: $1$ $ee$, $3$ $\mu \mu$ and $9$ $e \mu$ when
the expected numbers, scaled to $13$, are $3.3 \pm 0.5$, 
$2.8 \pm 0.5$ and $6.8 \pm 0.8$~\cite{CDFdil2}.
An excess in the $e \mu$ channel and a deficit in the $ee$ channel
have been found in both Run 1 and Run 2 counting experiments. 
We believe that this asymmetry does not look easy to explain 
simply in terms of different efficiencies, acceptances 
and backgrounds for the three classes of events.

\begin{table}[t!]
\begin{center} 
\begin{tabular}{ c c c c } 
\hline\hline 
{\bf MODELS} & $\mathbf{Br(t \to c S^0)}$ & $\mathbf{Br(S^0 \to \mu \tau)}$ &   
$\mathbf{Br(t \to c \mu\tau)}$\\ 
\hline 
{\bf MSSM}  & $10^{-3} - 10^{-4}$ &  $10^{-4} - 10^{-5}$ 
 &  $10^{-7} - 10^{-9}$\\
\hline 
{\bf Flavon} & $1. \times 10^{-1}$  & $4.\times 10^{-1}$ &  $4.\times 10^{-2}$ \\
\hline 
{\bf THDM-III}  & $10^{-1}$  & $2\times 10^{-5}$  &  $2 \times 10^{-6}$  \\
\hline 
{\bf Strong Dynamics}  & --  &  --  &  $0.1 \div 0.2$  \\
\hline\hline 
\end{tabular} 
\end{center} 
\vspace{-0.55cm}
\caption{
Branching ratios for $t \to c S^0$,  $S^0\to \tau \mu$ 
and for $\,\,\,\,t \to c \mu \tau$. In models with strong 
dynamics, DFV decays proceed through a direct contact interaction.} 
\vspace{-0.4cm}
\end{table}

\vspace{-0.7cm} 
\subsection{B.   $\,\,$Model  $\,$Independent  $\,$Parametrization} 
\vspace{-0.33cm} 

In the context of the previously discussed theoretical scenarios, MTV top decay
$t \rightarrow c \mu \tau$ are available with branching ratios up to $0.2$.
$Br(t\rightarrow u \tau \mu)$ is generally suppressed, except for 
the case of strongly interacting model, where both the 
branching ratios may reach similar values.
If $\mathcal{B}$ is the branching ratio for MTV top decays:
$\mathcal{B} \equiv Br(t \rightarrow c(u) \mu \tau)$ 
and we assume no other significatively accessible decay channels, either than the 
SM and the MTV itself: 
$Br(t \rightarrow W b)= 1 - \mathcal{B}$.
The possible decay modes for a $t \bar{t}$ pair are then:
{\it i)} purely SM decays occuring with a branching ratio of $(1- \mathcal{B})^2$; 
{\it ii)} mixed SM-MTV decays accessible with a branching ratio of $2{\mathcal{B}}(1-{\mathcal{B}})$;
{\it iii} purely MTV top decays having a branching ratio of: $\mathcal{B}^2$.
In total $(1- \mathcal{B})^2 + 2 \mathcal{B} (1-\mathcal{B}) + \mathcal{B}^2=1$.
Then SM-MTV $t\bar{t}$ decays, with one leg decaying into $bW_{\ell}$ and 
the other one into $c\mu \tau_{h}$, may fake the SM dilepton signature.

\noindent
We found that a $\mathcal{B} \simeq 0.10 \div 0.25$ 
is able to provide an excess in the $e \mu$ channel and a deficit in 
the number of $ee$ events that is consistent 
with CDF Run 1 and Run 2 data.
Anyhow, we believe that, at the present, dilepton analysis cannot rule out any of the 
model studied in present paper.

\vspace{-0.45cm} 
\section{V.   $\,$Conclusions  $\,$and  $\,$perspectives} 
\vspace{-0.35cm}

Triggered by CDF top dilepton analysis, that observed
an excess in the number of $e \mu$ and a deficit in the number 
of $ee$ top dilepton candidates, we studied the possibility 
that double flavor-violating top decays: $t\rightarrow c(u) \tau \mu$ 
may contaminate the top sample.
If MSSM predicts values for $Br(t\rightarrow c \tau \mu)$ 
up to about $10^{-7}$ and Flavor Models gives $\simeq  4 \times 10^{-2}$, 
Strongly Interacting Models offer the possibility  to reach 
$\simeq 0.10 \div 0.20$.
We found that such a value can explain the pattern observed in the data.
As at present, top dilepton analysis cannot rule out any of the 
model discussed above, further studies will be needed in order
to determine whether such new physics effects will be confirmed 
at Tevatron or may show up at lower branching ratios in 
next coming LHC experiments.  

\vspace{-0.35cm} 
\section{Acknowledgments}  
\vspace{-0.5cm} 
This work was supported by NSF and CONACYT-SNI (Mexico).

 


\begin{thebibliography}{99}
\bibitem{backgrounds}
Main backgrounds are Drell-Yan 
($Z^{*}/\gamma \rightarrow ee$, $\mu \mu$), $Z \rightarrow \tau \tau$ and
$WW$ production. Other sources are processes with a real lepton and a 
jet or a track faking a second lepton and processes in which mismeasured 
muon tracks can result in an overstimate of the $\not\!\!\!E_{\rm T}$.
\end{thebibliography}
\end{document}